\documentclass[
 ,secnumarabic%
 ,twocolumn%
,amssymb,
nobibnotes,
nofootinbib,
aps, prl,
showpacs,showkeys]{revtex4}
\usepackage{docs}%
\usepackage{bm}%
\usepackage{graphicx}
\expandafter\ifx\csname package@font\endcsname\relax\else
 \expandafter\expandafter
 \expandafter\usepackage
 \expandafter\expandafter
 \expandafter{\csname package@font\endcsname}%
\fi



\begin{document}

\title{$C_5^A$ axial form factor from  bubble chamber experiments}

\begin{abstract}
A careful reanalysis of both Argonne National Laboratory and Brookhaven National Laboratory data for weak single pion production is done. We consider deuteron nuclear effects and normalization (flux) uncertainties in both experiments. We demonstrate that these two sets of data are in good agreement. For the dipole parametrization of $C_5^A(Q^2)$, we obtain $C_5^A(0)=1.19\pm 0.08$, $M_A=0.94\pm 0.03$~GeV. As an application we present the discussion of the uncertainty of the neutral current  1$\pi^0$ production cross section, important for the T2K neutrino oscillation experiment.
\end{abstract}

\author{K. M. Graczyk$^{\dag}$}
\email{kgraczyk@ift.uni.wroc.pl}
\author{D. Kie\l czewska$^{\ddag,\diamondsuit}$}
\author{P. Przew\l ocki$^\diamondsuit$}
\author{J. T. Sobczyk$^\dag$}

\affiliation{$^\dag$Institute of Theoretical Physics, Wroc\l aw University,
 pl. M. Borna 9, 50-204, Wroc\l aw, Poland\\
$^\ddag$ Institute of Experimental Physics, University of Warsaw, Hoza 69,
00-681 Warsaw, Poland\\
$^\diamondsuit$ A. Soltan Institute for Nuclear Studies, Hoza 69, 00-681 Warsaw, Poland }

\maketitle

\section{Introduction}

The problem of the correct evaluation of neutrino single pion production (SPP) cross sections is  interesting by itself and also important for future neutrino oscillation experiments.
Recent measurements suggest that SPP cross section may be 20\%-25\% larger than what was assumed in the past \cite{MBCCPi+}. Understanding of nuclear effects in neutrino interactions in the 1 GeV energy region is still not satisfactory.  In order to cope with the deficit of events in the low $Q^2$ region experimental groups have introduced effective quantities like $\kappa$ in charge current quasielastic scattering (CCQE) \cite{MBkappa} or very large value of Fermi momentum in Pauli blocking \cite{MinosPauli blocking}. These difficulties have the positive effect that neutrino cross section experimental results are more often presented in the form of raw data, which are only efficiency corrected. Such measurements are free from dependence on the models implemented in the Monte Carlo generators of events but extraction of free nucleon cross-sections requires good knowledge of final state interactions (FSI) effects. For example, in the recent MiniBooNE neutral current (NC) 1$\pi^0$ cross-section measurement on $CH_2$ there were large carbon pion
absorption and charge exchange nuclear effects \cite{MBNC1Pi0}.

For this reason it makes sense to come back to old but relatively good statistics SPP data on deuteron obtained in bubble chamber experiments in the Argonne National Laboratory (ANL) \cite{Radecky:1981fn} and in the
Brookhaven National Laboratory (BNL) \cite{Kitagaki:1990vs}. The big advantage of these measurements is that deuteron nuclear effects are  easier to control.  It was claimed that SPP data from two experiments
are in disagreement \cite{Lalakulich:2005cs, Wascko:2006tx}, because of different total cross-sections and shapes of $d\sigma/dQ^2$.
Apparently it is true that the total cross-sections reported by the BNL experiment seem to be systematically larger than those obtained by the ANL. But it should be noted that in both experiments there are large normalization uncertainties of the measured cross-section due to imprecise knowledge of the neutrino flux. It is therefore desirable to perform a careful simultaneous reanalysis of two sets of data and this is the aim of our paper. We notice that during the last few years several theoretical models  have been proposed \cite{Lalakulich:2005cs, Sato:2003rq, Hernandez:2007qq, Leitner:2006ww} to describe the weak pion production off nucleon/nucleus. They were fine-tuned to the neutrino scattering data (usually only to the ANL data) not corrected to the effects which we consider in this paper.

The details of our approach will be given in the next sections of the paper, and here we would like to outline the main points:
\begin{enumerate}
\item We focus on the $\nu_\mu p\to \mu^- p \pi^+$ reaction  because for this process it seems to be reasonable to neglect a small nonresonant contribution. However, we notice that some of the current theoretical approaches include nonresonant background also for this channel (see \cite{Hernandez:2007qq}).

\item We will include deuteron nuclear effects applying the approach from Refs. \cite{AlvarezRuso:1998hi, Singh:2003a}.

\item In our statistical analysis we use the $\chi^2$ function with a contribution also coming  from the overall normalization (flux) uncertainty, and this turns out to be  important.
\end{enumerate}

Our main result is that ANL and BNL data are in quite a good agreement. We demonstrate this first by providing the value of $\chi^2$ for our best fits but then also using more sophisticate methods of checking the self-consistency of the two independent sets of data.

We present our results in the form of a fit for the $C_5^A(Q^2)$ form factor. We investigated several parametrizations with two options: either keeping $C_5^A(0)$ as a free parameter or fixing its value using the argument motivated by the partially conserved axial current (PCAC) hypothesis. In the most standard dipole parametrization with the $C_5^A(0)$ kept as a free parameter and with correction from deuteron effects included our best fit is: $C_5^A(0)=1.19\pm 0.08$, $M_A=0.94\pm 0.03$~GeV.

We also present two applications for our results. The first one is motivated by a need to evaluate the NC 1$\pi^0$ cross-section in the T2K experiment \cite{T2KLOI, Terri:2009zz}. We will estimate the uncertainties of the NC 1$\pi^0$ cross-section. In order to make the evaluation realistic, we will use two different Monte Carlo event generators with all the nuclear effects included. The second application is a comparison with recent
measurements of the CC1$\pi^+$/CCQE ratio done by MiniBooNE on the CH$_2$ target \cite{MBRatio}. This measurement is particulary important because it is free from the flux normalization uncertainty. We obtain good agreement with the data.

Our paper is organized as follows. Sec. 2 contains basic definitions and notation, and in  Sec. 3 we introduce various parametrizations of the axial form factors. In Sect. 4 we describe our approach to treat deuteron nuclear effects.  Sec. 5 is the most important, as we present the methodology of our analysis: we define $\chi^2$ and for completeness we provide all the necessary information about the form in which ANL and BNL data are given and used in our numerical analysis. In Sec. 6 we show our results for various parameterizatons of the axial form factor. The complete set of results is given in  Tables. \ref{table_dipole_results} and \ref{table_adler_results}. Sec. 6 contains also detailed discussion of our results as well as a formal demonstration that ANL and BNL data sets are self-consistent. In Sec. 7 applications of our results are given and  last Sec. 8 contains our final remarks.

\section{Adler-Rarita-Schwinger formalism}

We analyze the experimental data for the charged current (CC) neutrino-proton reaction:
\begin{equation}
\label{process_CC} \nu(k) + p(p) \to \mu^-(k') + \Delta^{++}(p'),
\end{equation}
where $k$, $k'$, $p$ and
$p'$ denote neutrino, muon, proton and $\Delta(1232)$ resonance four-momenta.
The four-momentum transfer and its square are
\begin{equation}
q = p'-p = k-k',\qquad Q^2\equiv -q^2,
\end{equation}
and the hadronic invariant mass is
\begin{equation}
W^2  = {p'}^2 = (p+q)^2.
\end{equation}

One way to describe the reaction (\ref{process_CC}) is to
apply the Adler-Rarita-Schwinger formalism. The final hadronic state is a
3/2-spin resonance described as a Rarita-Schwinger field. The
transition from the nucleon to $\Delta^{++}$ is given as a
matrix element of the weak hadronic current, which has the standard vector-axial structure
\cite{Berman:1965iu,Adler,Adler:1975mt,Llewellyn Smith:1971zm}:
\begin{equation}
\mathcal{J}^{CC}_\mu = \mathcal{J}^{V}_\mu + \mathcal{J}^A_\mu.
\end{equation}
Both vector and axial parts are expressed in terms of
several form factors.

Under general assumptions, the vector part can be expressed by means of three form factors,
$C_3^V(Q^2)$, $C_4^V(Q^2)$, and $C_5^V(Q^2)$:
\begin{widetext}
\begin{equation}
\left<\Delta^{++}(p') \right| \mathcal{J}_\mu^{V} \left| N(p)
\right> = \sqrt{3} \bar{\Psi}_\lambda (p') \left[ g^{\lambda}_{\
\mu} \left( \frac{C_3^V}{M}\gamma_\nu + \frac{C_4^V}{M^2}{p'}_\nu +
\frac{C_5^V}{M^2}p_\nu\right) q^\nu - q^\lambda \left(\frac{C_3^V}{M}\gamma_\mu + \frac{C_4^V}{M^2}p'_\mu +
\frac{C_5^V}{M^2}p_\mu\right) \right]\gamma_5 u(p),
\end{equation}
\end{widetext}
where $M$ is the nucleon mass,
$\Psi_\mu (p')$ is the Rarita-Schwinger field, and $u(p)$ is the Dirac
spinor.

The axial part depends on four form factors $C_3^A(Q^2)$,
$C_4^A(Q^2)$, $C_5^A(Q^2)$ and $C_6^A(Q^2)$ \cite{Jones:1972ky}:
\begin{widetext}
\begin{equation}
\left<\Delta^{++} (p')\right|\mathcal{J}_{\mu}^{A} \left|N(p)\right>
= \sqrt{3}\bar{\Psi}_\lambda (p') \left[ g^{\lambda}_{\
\mu}\left(\gamma_\nu \frac{C_3^A}{M} + \frac{C_4^A}{M^2}{p'}_\nu
\right)q^\nu- q^\lambda\left(\frac{C_3^A}{M}\gamma_\mu +
\frac{C_4^A}{M^2}{p'}_\mu\right) + g^{\lambda}_{\ \mu}C_5^A +
\frac{q^\lambda q_\mu}{M^2} C_6^A\right]u(p).
\end{equation}
\end{widetext}

The differential cross-section for the reaction (\ref{process_CC})
reads
\begin{equation}
\sigma_{th} (E, Q^2, W)\equiv\frac{d^2 \sigma}{dW dQ^2} =
\frac{\widetilde{G}^2 W}{64 \pi^2 M E^2  }L^{\mu\nu}W_{\mu\nu},
\end{equation}
where $\widetilde{G} = G \cos\theta_c$, $G$ is the Fermi
constant (for a neutral current reaction $\widetilde{G} = G$), and $\theta_c$ is the  Cabibbo angle. The neutrino energy is denoted by $E$.

$W_{\mu\nu}$ and $L_{\mu\nu}$ are hadronic and leptonic tensors
defined as:
\begin{widetext}
\begin{eqnarray}
\label{tensor_W_delta} W_{\mu\nu} &=&  \frac{1}{4M
M_\Delta}\frac{1}{2}\sum_{spin}
\left<\Delta^{++},p'\right|{\mathcal{J}_\mu^{CC}}\left| p\right>
\left<\Delta^{++},p'\right|{\mathcal{J}_\nu^{CC}}\left| p\right>^*
 \frac{\Gamma_\Delta/2}{((W - M_\Delta)^2 + \Gamma_\Delta^2/4)} , \\
{\rm L}_{\mu\nu} &=& 8\left({k'}_{\mu}k_\nu + k_{\mu}{k'}_\nu
-g_{\mu\nu}{k'}_\alpha k^\alpha \mp
i\epsilon_{\mu\nu\alpha\beta}k^\alpha {k'}^\beta  \right).
\end{eqnarray}
\end{widetext}

$\Gamma_\Delta(W)$ is the $\Delta$ width, for which we assume the $P$-wave ($l=1$) expression
\begin{equation}
\label{szerokosc_polowkowa} \Gamma_\Delta = \Gamma_0
\left(\frac{q_{cm}(W)}{q_{cm}(M_\Delta)}\right)^{2l +1
}\frac{M_\Delta}{W}
\end{equation}
with
\begin{equation}
\label{q_cm}  q_{cm}(W)=
\sqrt{\left(\frac{W^2+M^2-m_\pi^2}{2W}\right)^2 -  M^2}
\end{equation}
$M_\Delta=1232$~MeV, $m_\pi=139.57$~MeV is the charged pion mass.

\subsection{Neutral current scattering}

We introduce the amplitude for the $p \to \Delta^{++}$ weak
transition:
\begin{equation}
\label{Amplituda_Delta}
\mathcal{A}_{\Delta} = \mathcal{A}\left( p\to \Delta^{++}\to p \pi^+ \right)
\end{equation}

Then the amplitudes for the other two channels for CC neutrino-neutron scattering are expressed in terms of the appropriate  Clebsch-Gordan coefficients:
\begin{equation}
-\frac{\sqrt{2}}{3}\mathcal{A}_{\Delta} = \mathcal{A}\left( n\to \Delta^{+}\to p \pi^0 \right)
\end{equation}
\begin{equation}
\frac{1}{3}\mathcal{A}_{\Delta} =  \mathcal{A}\left( n\to \Delta^{+} \to n \pi^+\right)
\end{equation}

A way to describe SPP in NC neutrino-nucleon  scattering was proposed in Ref. \cite{Fogli_and_Narduli}. First of all, the charge current and neutral current amplitudes are related through the Clebsch-Gordan relations:
\begin{eqnarray}
\mathcal{A}'(\nu p \to \nu \Delta^{+} \to \nu \pi^0 p ) &=& \frac{\sqrt{2}}{3}{\mathcal{A}'}_{\Delta} \\
\mathcal{A}'(\nu n \to \nu \Delta^{0} \to \nu \pi^0 n ) &=& \frac{\sqrt{2}}{3}{\mathcal{A}'}_{\Delta} \\
\mathcal{A}'(\nu p \to \nu \Delta^{0} \to \nu \pi^+ n ) &=& \frac{1}{3}{\mathcal{A}'}_{\Delta} \\
\mathcal{A}'(\nu n \to \nu \Delta^{+} \to \nu \pi^- p ) &=& \frac{1}{3}{\mathcal{A}'}_{\Delta}.
\end{eqnarray}
Additionally, the amplitude ${\mathcal{A}'}_{\Delta}$ is computed like (\ref{Amplituda_Delta}), but as an input one needs to apply hadronic current according to the recipe from the standard model:
the vector part is multiplied by  $(1 - \sin^2\theta_W)$, $\theta_W$ is the Weinberg angle, and the axial part is left unchanged
\begin{equation}
{\mathcal{A}'}_{\Delta} = {\mathcal{A}}_{\Delta}\bigl( C_i^V \to (1 - \sin^2\theta_W) C_i^V ,C_i^A \to C_i^A\bigr).
\end{equation}

\section{Form factors \label{section_form_factors} }

\subsection{Vector form factors \label{subsection_vector_form_factors}}

In an older analysis (see e.g. \cite{Schreiner:1973mj}) an
additional constraint on the vector form factors was imposed:
\begin{equation}
\label{su6_cv_limit}
C_5^V(Q^2) = 0, \quad C_4^V(Q^2) = - \frac{M}{W}C_3^V(Q^2)
\end{equation}
as motivated by the quark model relations [$SU(6)$ symmetry
relation, for details see e.g. Ref. \cite{Liu:1995bu}].
The above relations describe the dominance of the magnetic amplitude $M1$, while
the electric amplitude $E2$ vanishes. With constraints given in (\ref{su6_cv_limit}), the vector current is expressed
by only one unknown function, which can be extracted from the electroproduction data.

In the original analysis of the ANL and BNL data two parametrizations of the $C_3^V(Q^2)$ were considered. The first one proposed by Dufner and Tsai \cite{Dufner:1967yj} (applied to ANL data analysis \cite{Barish:1978pj}) reads
\begin{equation}
\label{cv_old1}
C_3^V(Q^2) = 2.05 \sqrt{1 + 9 Q } e^{-3.15 Q }, \quad Q =
\sqrt{Q^2},
\end{equation}
and the other one is of a simple dipole form,
\begin{equation}
\label{cv_old2}
C_3^V(Q^2) = \frac{2.05}{\displaystyle \left(1  + \frac{Q^2}{0.54 \,
\mathrm{GeV}^2}\right)^2}.
\end{equation}

In our analysis we will use instead the recent experimental fits proposed in
\cite{Lalakulich:2006sw}:
\begin{eqnarray}
C_3^V(Q^2) &=& 2.13 \left( 1 +\frac{Q^2}{4
M_V^2}\right)^{-1}G_D(Q^2), \\
C_4^V(Q^2) &=& -1.51 \left( 1 +\frac{Q^2}{4
M_V^2}\right)^{-1}G_D(Q^2), \\
\label{cv_new}
C_5^V(Q^2) &=&0.48\left( 1 +\frac{Q^2}{0.776 M_V^2}\right)^{-1}G_D(Q^2),
\end{eqnarray}
where
\begin{equation}
G_D(Q^2) = \left(1+\frac{Q^2}{M_V^2} \right)^{-2}, \quad
\mathrm{and}\quad M_V =0.84~\mathrm{GeV}.
\end{equation}

It is clear that the use of different vector form factor has an
impact on the axial contribution as obtained from a fitting
procedure to the neutrino scattering data. We will return to this
point in the discussion.
\begin{figure*}[t!]
\centering{
\includegraphics[width=1.0\textwidth, height=10.5cm]{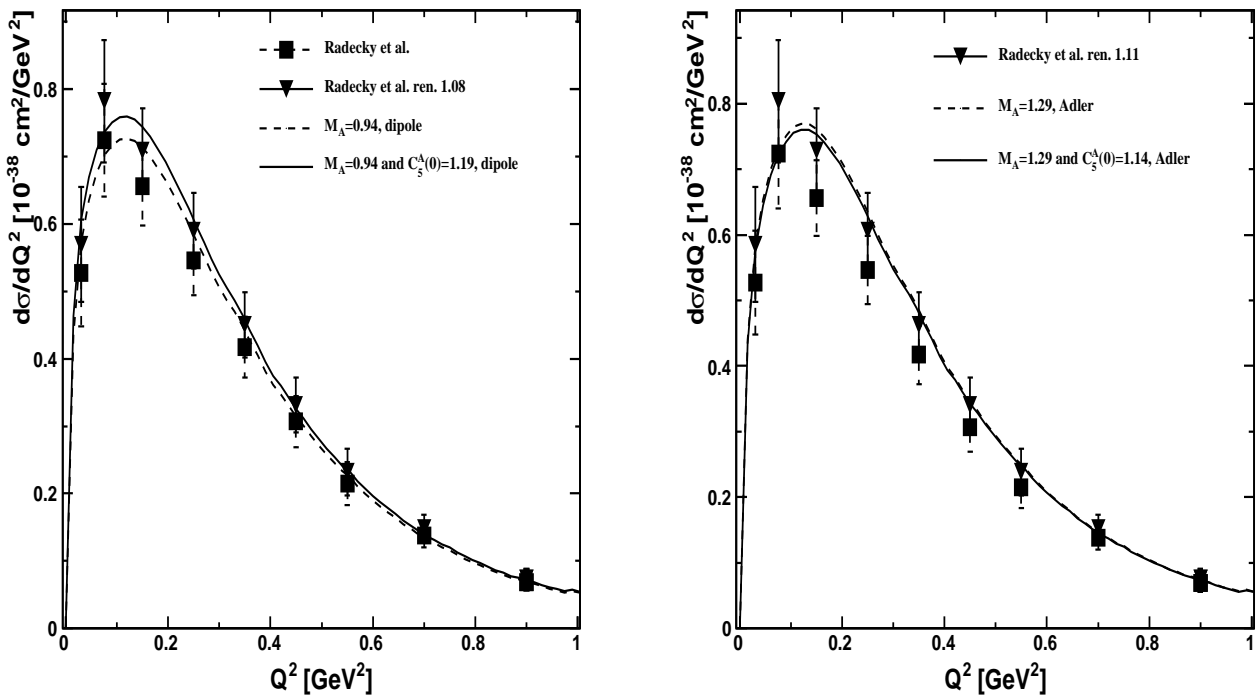}}
\caption{
Differential cross-sections for $\nu+ d \to \mu^- + n + p
+\pi^+$ scattering computed for the ANL beam. The solid/dashed
lines denotes the best fit obtained  with $C_5^A(0)$ fitted/fixed (with value 1.15). In the left panel the cross-sections computed with  the dipole functional form of $C_5^A(Q^2)$ (\ref{c5a_dipol}) are shown, while in the right panel results obtained with Adler parametrizations (\ref{c5a_adler}) are presented.  The black squares denote the experimental data  \cite{Radecky:1981fn}, while the black triangles denote the experimental data multiplied by
factor of 1.08 (left panel) and 1.11 (right panel)  -- the renormalization factors were obtained from the global fits (see Tables \ref{table_dipole_results} and \ref{table_adler_results}). The theoretical cross-sections are modified to include deuteron nuclear effects. The cut $W<1.4$ GeV on the hadronic invariant mass was imposed.
\label{fig_anl_q2}
}
\end{figure*}

\begin{figure*}[t!]
\centering{
\includegraphics[width=1.0\textwidth]{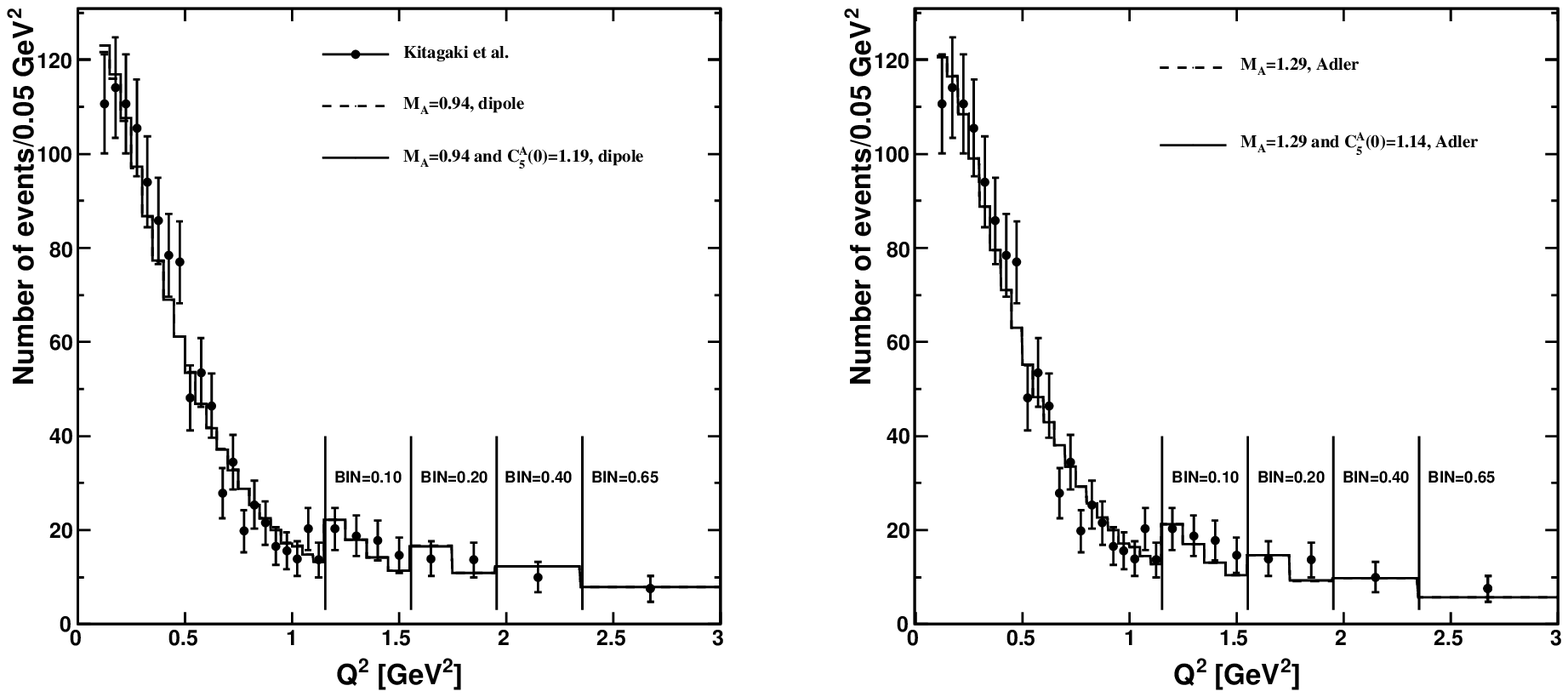}}
\caption{
Distribution of the number of events of the BNL experiment for
$\nu+ d \to \mu^- + n + p +\pi^+$ scattering.  The solid/dashed
lines denotes the best fit obtained  with $C_5^A(0)$ fitted/fixed (with a value of 1.15). In the left panel the cross-sections computed with  the dipole functional form of $C_5^A(Q^2)$ (\ref{c5a_dipol}) are shown, while in the right panel results obtained with Adler parametrizations (\ref{c5a_adler}) are presented.   The black dots denote the experimental data \cite{Kitagaki:1990vs}. The theoretical cross-sections are modified to include deuteron nuclear effects. The cut $W<1.4$ GeV on the hadronic invariant mass was imposed. For higher $Q^2$ (above 1.15~GeV) the data bins were joined, to get better statistics. The new $Q^2$ bins have the following widths: 0.1 for $Q^2 \in (1.15,1.55)$, 0.2~GeV$^2$ for $Q^2 \in (1.55, 1.95)~\mathrm{GeV}^2$, 0.4~GeV$^2$ for $Q^2 \in (1.95,2.35)~\mathrm{GeV}^2$, and 0.65~GeV$^2$ for  $Q^2 \in (2.35, 3)~\mathrm{GeV}^2$.
\label{fig_bnl_q2}}
\end{figure*}

\subsection{ Axial form factors\label{subsection_axial_form_factors}}

The main contribution to the axial current comes from $C_5^A(Q^2)$, which is an analog of $C_3^V(Q^2)$ in the vector current -- it describes the $(M1)^A$ amplitude. Similarly, $C_3^A(Q^2)$ is
an analog of $C_5^V (Q^2)$ because it contributes to the $(E2)^A$
amplitude \cite{Liu:1995bu}. In the
$SU(6)$ symmetry limit $(E2)^A=0$, $C_3^A(Q^2)=C_6^A(Q^2)=0$ and only $C_4^A(Q^2)$ and $C_5^A(Q^2)$ are nonvanishing.

The $SU(6)$ symmetry relations are only approximate, and they become broken by color
hyperfine interaction. A lot of effort was done to compute all axial
form factors directly from the quark models \cite{BarquillaCano:2007yk}
but the results do not reproduce the data sufficiently well.

In the phenomenological analysis there is no enough experimental data to extract
all the axial form factors separately. Therefore some extra constraints must be imposed:
\begin{description}
\item[(i)] typically one sets $ C_3^A(Q^2)=0$;
\item[(ii)] the Adler model \cite{Adler} suggests
\begin{equation}
C_4^A(Q^2) = - C_5^A(Q^2)/4;\label{C4_C5}
\end{equation}

\item[(iii)] the PCAC hypothesis implies
\begin{equation}
C_6^A(Q^2) =   \frac{M^2}{m_\pi^2 + Q^2} C_5^A(Q^2)
\end{equation}
with $m_\pi$ being the pion mass.
\end{description}

$C_5^A(0)$ can be
evaluated from the off-diagonal Goldberger-Treiman relation (for a review see e.g. \cite{Thomas_book}). The
updated value is reported in Ref. \cite{BarquillaCano:2007yk}:
\begin{equation}
\label{C5A_0_current} C_5^A(0) = \frac{g_{\pi N\Delta}  f_\pi
}{\sqrt{6} M} =  1.15 \pm 0.01,
\end{equation}
where the values $g_{\pi N\Delta}(q^2 = m_\pi^2 )=28.6 \pm 0.3$ and $f_\pi =
92.4$~MeV were used.

In the models presented in Refs. \cite{Hernandez:2007qq, BarquillaCano:2007yk, Hemmert:1994ky}, the value of $C_5^A(0)$ is
smaller than the one obtained from formula (\ref{C5A_0_current}). Therefore
it seems reasonable to use the experimental data in order to evaluate its value. In this paper we perform a comprehensive fit to the existing data on elementary interactions from ANL and BNL experiments.
The value of $C_5^A(0)$ was already studied
experimentally by Barish et al. \cite{Barish:1978pj} with the
conclusion that $C_5^A(0)=1.2$ is in agreement with the
data with an accuracy of 20\%.

In recent years the functional form of $C_5^A(Q^2)$ has been also a
subject of intensive research
\cite{Lalakulich:2005cs, Hernandez:2007qq, AlvarezRuso:1998hi, Adler, Schreiner:1973mj, Liu:1995bu, BarquillaCano:2007yk, Leitner:2008fg, Graczyk:2007bc, Jarek_nuint09}.
A possible parametrization of $C_5^A(Q^2)$ is  based on a comparison
of the Rarita-Schwinger formalism with the predictions of the Adler
model \cite{Schreiner:1973mj}:
\begin{equation}
\label{c5a_adler} C_i^A(Q^2) = C_i^A(0) \left(1+ \frac{a_i Q^2}{b_i+
Q^2} \right)\displaystyle \left(1 +\frac{Q^2}{M_A^2} \right)^{-2}
\end{equation}
with values:
\begin{eqnarray}
C_3^A(0)&=& a_3=b_3 =0, \quad C_4^A(0)= -C_5^A(0)/4, \nonumber \\
\quad C_5^A(0)&=& 1.2, \nonumber \\
\quad a_4 &=& a_5 = -1.21, \quad b_4 = b_5 = 2.
\end{eqnarray}

In our analysis the first ansatz for the functional form of $C_5^A(Q^2)$
will be as above. The fitted parameters will be either only $M_A$ or both $M_A$ and $C_5^A(0)$.

In the second fit we consider the simplest possible functional form, namely the dipole parametrization:
\begin{equation}
\label{c5a_dipol} C_5^A(Q^2) =
\frac{C_5^A(0)}{\displaystyle \left(1 + \frac{Q^2}{M_A^2}\right)^2},
\quad C_5^A(0)=1.15.
\end{equation}
Again, we will fit either only $M_A$ or both $M_A$ and $C_5^A(0)$.

Let us emphasize that in both cases when $C_5^A(0)$ was treated as a free parameter we kept the constraint (\ref{C4_C5}).

\section{Neutrino-deuteron scattering \label{neutrino_deuteron_scattering}}

In both the ANL and BNL experiments for most of the exposition the detectors were
filled with deuteron. The nuclear effects for the $\Delta^{++}$
production were discussed in Ref. \cite{AlvarezRuso:1998hi}. The final state interactions can be neglected and the effect comes from the fact that the target proton is bound.

In \cite{AlvarezRuso:1998hi} the deuteron wave functions were considered to be obtained in three
different nuclear models:   Hulthen \cite{Hulthen}, Paris
\cite{Lacombe:1981eg}, and Bonn \cite{Machleidt:1987hj}.
The results of \cite{AlvarezRuso:1998hi} indicate that the nuclear
effects for $\Delta^{++}$ production are
larger than for the quasielastic scattering, in which case the main modification is a reduction of the cross-section in the region
of small $Q^2<0.05$~GeV$^2$ due to Pauli blocking while nuclear effects are negligible for $Q^2>0.15$~GeV$^2$ \cite{Singh:1971md}.
According to the results of \cite{AlvarezRuso:1998hi} in the
case of $\Delta^{++}$ excitation the nuclear effects are slowly varying
with $Q^2$ and they reduce the differential cross-section in $Q^2$
by $\sim 10\%$ in the case of the ANL beam and $\sim 5\%$ for neutrinos
of energy $1.6$~GeV (a typical value for the BNL beam).

In our analysis we therefore assume that the
neutrino--deuteron $\Delta^{++}$ excitation  differential cross-section in $Q^2$ gets modified by a function $R(Q^2)$ with respect to the neutrino-proton cross-section:
\begin{equation}
\left( \frac{d\sigma}{dQ^2}\right)_{deuteron} = R(Q^2) \left(
\frac{d\sigma}{dQ^2}\right)_{free\; target}
\end{equation}
In general $R(Q^2)$ is a function of the neutrino energy $E$, but for the ANL
experiment we use the result of Ref. \cite{AlvarezRuso:1998hi}
(Fig. 5), where the flux averaged differential cross-section is
plotted with and without deuteron effects. Based on that we
extracted:
\begin{equation}
R_{ANL}(Q^2)  = \frac{\left( d \sigma(\nu d \to \mu^-
n\Delta^{++})/dQ^2\right)_{deuteron}}{\left(d\sigma(\nu  p \to
\mu^- \Delta^{++})/dQ^2\right)_{free\; target}}
\end{equation}

In the case of the BNL experiment we use $R(Q^2)$ evaluated at $E=1.6$~GeV.

There is an additional ambiguity in applying the results from \cite{AlvarezRuso:1998hi} because the calculated reduction of the cross-section depends
on the model of the deuterium potential.

We considered all three  nucleon-nucleon
potentials mentioned above. However, in this paper only results obtained
with the Paris potential are presented. The Bonn potential gives rise to very similar results.
With the Hulthen  potential, the results are still comparable, but the impact of nuclear effects is smaller.

In \cite{Campbell} early ANL data on mostly hydrogen
targets i.e. without nuclear effects, are presented. There are
altogether 153 such events out of which it is estimated that 105 are on the hydrogen target. This sample of events was
used in the analysis of Schreiner and Von Hippel \cite{Schreiner:1973mj}.

\section{reanalysis of the bubble chamber deuterium data}

We look for a simultaneous fit to both ANL and BNL data by applying the $\chi^2$ method. The  best
fit is obtained by minimizing a function:
\begin{equation}
\label{chi2_total} \chi^2 = \chi^2_{ANL} + \chi^2_{BNL},
\end{equation}
where $\chi^2_{ANL}$ and $\chi^2_{BNL}$ are defined for each data set separately.

In both cases the  $\chi^2$ is given by the standard formula with an
additional quadratic term which comes from the total systematic uncertainty for the flux \cite{D'Agostini:1995fv}:
\begin{widetext}
\begin{equation}
\label{chi2_Ni} \chi^2_{}=\sum_{i=1}^{n_{}}
\left(\frac{N_{th,i}^{} - N_{i}^{}}{\Delta N_i^{}}
\right)^2 + \left(\frac{\displaystyle
\frac{\sigma_{tot-th}^{}}{\sigma_{tot-ex}^{}}\cdot
\frac{{N}_{exp}}{{N}_{}^{th}} -1}{ r_{} }\right)^2,
\end{equation}
or equivalently by
\begin{equation}
\label{chi2_sigma} \chi^2_{}=\sum_{i=1}^{n_{}}
\left(\frac{\sigma_{th}^{diff}(Q^2_i) - p
\sigma_{ex}^{diff}(Q^2_i)}{ p \Delta \sigma^{}_i} \right)^2 +
\left(\frac{p -1}{ r_{} }\right)^2,
\end{equation}
\end{widetext}
with:
\begin{equation}
p\equiv \frac{ \sigma_{tot-th}^{}} { \sigma_{tot-exp}^{} }
\frac{ N_{}^{exp}  } { N_{}^{th}  },
\end{equation}
$N_i^{}$ and $N_{th,i}^{}$ are experimental results and theoretical predictions for
the number of events in the $i$-th $Q^2$-bin, $\Delta N_i^{}$
is the experimental result uncertainty (it is the sum of
statistical and systematic uncorrelated contributions),  $\sigma_{tot-exp}^{}$ and  $\sigma_{tot-th}^{}$
are the experimental and theoretical flux averaged cross-sections measured and
calculated with the same cuts, and finally
\begin{equation}
N_{}=\sum_j N_j^{}, \qquad N_{}^{th}=\sum_j N_{th,j}^{}.
\end{equation}

In both above cases the form factor parameters are fitted but in the first case [Eq. (\ref{chi2_Ni})]  the overall number of events $N_{th}$ is fitted, while
in the second case (Eq. (\ref{chi2_sigma})) the fitting is applied
to $p$. In the presentation of our results, in both cases the final results are presented by giving the values of $p$.

\begin{figure}[t!]
\centering{\includegraphics[width=0.5\textwidth]{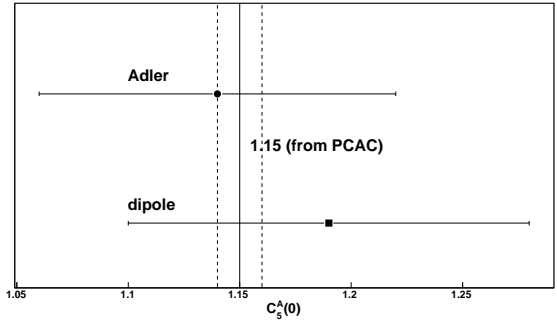}} \caption{
$C_5^A(0)$ fits and $1\sigma$ error bounds obtained by assuming dipole [Eq. (\ref{c5a_dipol})]
and Adler [Eq. (\ref{c5a_adler}] parametrizations of axial form
factors. The solid perpendicular line denotes the PCAC value quoted in
\cite{BarquillaCano:2007yk} together with $1\sigma$ error bounds. \label{fig_c5a}}
\end{figure}

\subsection{ANL data}
The ANL data are given in the form of normalized $ d\sigma/d Q^2 $ \cite{Radecky:1981fn} with a cut $W<1.4$~GeV. Information about the ANL
neutrino flux is provided in Ref. \cite{Barish:1977qk} (see Fig. 8). The uncertainties in the differential cross-section (both
statistical and uncorrelated systematic) are given in the original paper.

In the case of  ANL experiment the data consists of 1115 (871) corrected (raw) events. We use the
flux averaged
\begin{equation}
\frac{d\sigma  }{ d Q^2} (Q^2_i) \equiv \sigma_{exp}^{ANL,diff}(Q^2_i)
\end{equation}
data points (all are taken from Radecky \textit{et al.} \cite{Radecky:1981fn}) with uncertainties $\Delta\left(
{\sigma_i^{ANL,diff}}\right)$. All the numbers are given in table IV of Ref. \cite{Radecky:1981fn}. Cuts are imposed: for neutrino energy $E\in (0.5,6)$~GeV and for the hadronic
invariant mass $W<1.4$~GeV. The data covers the range in $Q^2$ from $Q^2=0.01$~GeV$^2$ to $Q^2=1$~GeV$^2$

The experimentally measured (flux averaged and with the cut $W<1.4$~GeV) cross-section is calculated to be:
\begin{equation}
\sigma_{tot-ex}^{ANL} = \sum_{i=1}^{n_{ANL}}\Delta Q_i^2\sigma_{exp}^{ANL,diff}(Q^2_i) = 0.31 \times 10^{-38} \mathrm{cm}^2,
\end{equation}
where $\Delta Q_i^2$ are the bin widths, $n_{ANL}=9$.

The theoretical formula for the differential cross-section in a given $Q^2$ bin is
the following:
\begin{widetext}
\begin{equation}
\sigma_{th}^{ANL,diff}(Q^2_i) =  \frac{1}{\Psi_{ANL}}\cdot\frac{1}{\Delta Q^2_i}\int_{Q^2_i - \Delta Q^2_i/2}^{Q^2_i + \Delta Q^2_i/2} d Q^2 \int_{E_{min}}^{E_{max}} d E \int_{M+m_\pi}^{1.4~GeV}dW\Phi_{ANL}(E) \sigma_{th}(E,Q^2,W),
\end{equation}
\end{widetext}
with
\begin{equation}
\Psi_{ANL} = \int_{E_{min}}^{E_{max}} dE \Phi_{ANL}(E),
\end{equation}
$E_{min}=0.5$~GeV and $E_{max}=6.0$~GeV.


The systematic uncertainty of the total cross-section resulting from an imprecise determination of the neutrino flux
is quoted in Ref. \cite{Radecky:1981fn} to be 15\% for $E\in (0.5, 1.5)$~GeV  and 25\% for $E>1.5$~GeV.
The flux was calculated on the basis of pion production cross-sections measured by Cho et al. \cite{Cho:1971}
for the same proton beam in a separate experiment. The calculation is described in Ref. \cite{Barish:1977qk} where
the neutrino flux computed based on pion production was compared with the flux  derived from
measurements of QE interactions. The conclusion was  that the latter flux was smaller by 21\%.
Therefore in our discussion, we assume the average
overall normalization uncertainty to be 20\%.
\begin{equation}
r_{ANL} = 0.20.
\end{equation}
In the case of ANL data it is natural to use 
formula  (\ref{chi2_sigma}) for $\chi^2$.

\begin{table*}[t!]
\centering{
\begin{tabular}{|c|c|c|c|c|c|c|c|}
\hline
 & & & & & & \\
 &  $M_A$ (GeV) &$C_5^A(0)$ & $p_{ANL}$ & $p_{BNL}$ & $\chi^2/NDF$ & GoF \\
 & & & & & & \\
  \hline
 & & & & & & \\
dipole, only $M_A$, free target  & $0.95\pm 0.04$ & & $1.15 \pm 0.06$ & $0.98 \pm 0.03$  & 25.5/28 & 0.60 \\
 & & & & & & \\
 dipole, only $M_A$, deuteron & $0.94\pm 0.04$ & & $1.04 \pm 0.06$  & $0.97 \pm 0.03$ & 24.5/28 & 0.65 \\
 & & & & & & \\
 & & & & & & \\
 dipole, $M_A$ and $C_5^A(0)$, free target  & $0.95\pm 0.04$ & $1.14\pm 0.08$  &  $1.15 \pm 0.11$ & $0.98 \pm 0.03$ & 25.5/27 & 0.54\\
 & & & & & & \\
 dipole, $M_A$ and $C_5^A(0)$, deuteron  & $0.94\pm 0.03$ & $1.19\pm 0.08$  &  $1.08 \pm 0.10$ & $0.98 \pm 0.03$ & 24.3/27 &  0.60\\
 & & & & & & \\
 \hline
\end{tabular}
} \caption{ The results obtained for fitting  the dipole
parametrization [Eq. (\ref{c5a_dipol})] of the axial form factor.  \label{table_dipole_results}}
\end{table*}
\begin{table*}[t!]
\centering{
\begin{tabular}{|c|c|c|c|c|c|c|c|}
\hline
 & & & & & & \\
 &  $M_A$ (GeV) &$C_5^A(0)$ & $p_{ANL}$ & $p_{BNL}$ & $\chi^2/NDF$ & GoF \\
 & & & & & & \\
  \hline
 & & & & & & \\
Adler, only $M_A$, free target  & $1.31\pm 0.06$ & & $1.23 \pm 0.06$ & $0.98 \pm 0.03$  & 27.1/28 & 0.50 \\
 & & & & & & \\
Adler, only $M_A$, deuteron   & $1.29\pm 0.07 $ & & $1.11 \pm 0.06$  & $0.98 \pm 0.03$ & 24.8/28 & 0.64 \\
 & & & & & & \\
 & & & & & & \\
 Adler, $M_A$ and $C_5^A(0)$, free target  & $1.31\pm 0.07$ & $1.1\pm 0.08$ &  $1.18 \pm 0.1$ & $0.98 \pm 0.03$ & 26.7/27 & 0.48\\
 & & & & & & \\
 Adler, $M_A$ and $C_5^A(0)$, deuteron  & $1.29\pm 0.07 $ & $1.14 \pm 0.08$ & $1.11 \pm 0.10$  & $0.98 \pm 0.03$ & 24.8/27 & 0.58 \\
 & & & & & & \\
 \hline
\end{tabular}
} \caption{ The results obtained for fitting  the Adler
parametrization [Eq. (\ref{c5a_adler})] of the axial form factor.  \label{table_adler_results}}
\end{table*}

\subsection{BNL data}

The BNL data are given in the form of the distribution of events in $Q^2$ \cite{Kitagaki:1986ct} with a cut $W<1.4$~GeV. Information about the
flux is given in \cite{Baker:1981su}  (Fig. 7). Reconstruction of the neutrino spectrum is also presented  in a later reanalysis done by Furuno \textit{et al.} \cite{Furuno:2003ng}. In our analysis we use the flux from this latest paper (Fig. 1, right panel). The overall
normalization (cross-section) is also provided but without cuts on the hadronic mass \cite{Kitagaki:1986ct}.
The beams in the ANL and BNL experiments are quite different so that the results are expected to give independent information on ${d\sigma}/{d Q^2}$.

The BNL data sample consists of 1803 (1610) corrected (raw) events
\cite{Kitagaki:1986ct}. The statistics is better than in the ANL
experiment by a factor of $\sim$ 50\% . In the case of the BNL
experiment we consider $dN/dQ^2$ taken from Fig. 10 of Ref.
\cite{Kitagaki:1986ct} with only statistical errors $\Delta N_i =
\sqrt{N_i}$. The events are collected under the condition
$W<1.4$~GeV and $E\in (0.5, 6)$~GeV for $Q^2\in (0,3)$ GeV$^2$. For $Q^2>1.2$~GeV$^2$ we
decided to combine some bins in order to get better statistics
(as illustrated in Fig. \ref{fig_bnl_q2}). We can do this because
in the analysis only the statistical errors of the experimental points
are taken into account.

The neutrino flux presented in the BNL papers \cite{Baker:1981su,Furuno:2003ng} was determined from measurements
of QE events \cite{Baker:1981su}. The QE axial mass derived only from the shape of $Q^2$ distribution with precision of 6\% was used
to calculate the total cross-section, and the observed rate of QE events allowed one to then calculate the neutrino flux
 with an uncertainty of less than 10\%.
The axial mass was later recalculated in Ref. \cite{Budd:2003} with new electro-magnetic form-factors and the updated value was found to be only 2\% smaller. The calculation of the neutrino flux coming from the same proton beam of 29~GeV/c from Alternating Gradient Synchrotron (AGS) was
presented in a paper by Ahrens et al.~\cite{Ahrens:1986ke} and
compared with the flux derived from QE events in a larger scintillator detector, which could measure the beam profile, contrary to the deuterium bubble chamber. It has been found that both fluxes differ by 10\% and the spectrum shapes agree very well.

Thus in our discussion below, we assume the normalization uncertainty of 10\% and in the main analysis we take
\begin{equation}
r_{BNL} = 0.10
\end{equation}
(In Sec. 6.2 we will consider the impact of different normalization uncertainties on the final results.)
To fit the normalization parameter for the BNL data we use the total
cross-section data published in  \cite{Kitagaki:1986ct}. For the
total cross-section data the hadronic invariant mass is
unconstrained, and the neutrino energy range is $E\in (0.5, 3)$~GeV.

In its analysis, the BNL collaboration  used only the events satisfying $Q^2>0.1$~GeV$^2$. Two different justifications for that can be found. In
\cite{Kitagaki:1990vs} it is written that the efficiency of reconstructing events for $Q^2<0.1$~GeV$^2$ is very low. This point
is later  investigated in detail and the dependence of the reconstructed parameters (axial mass) on $Q^2_{min}$ is shown. The value of
the fit to the axial mass is rather stable for $Q^2_{min}>0.06$~GeV$^2$. In \cite{Kitagaki:1990vs} one can also find
 a justification that nuclear effects play an important role in the $Q^2<0.05$~GeV$^2$ region.

In the reanalysis of the BNL data done in Ref. \cite{Furuno:2003ng} it is stressed that $\frac{dN}{dQ^2}$ data are presented with no nuclear corrections.

\begin{figure*}[t!]
\centering{
\includegraphics[width=\textwidth]{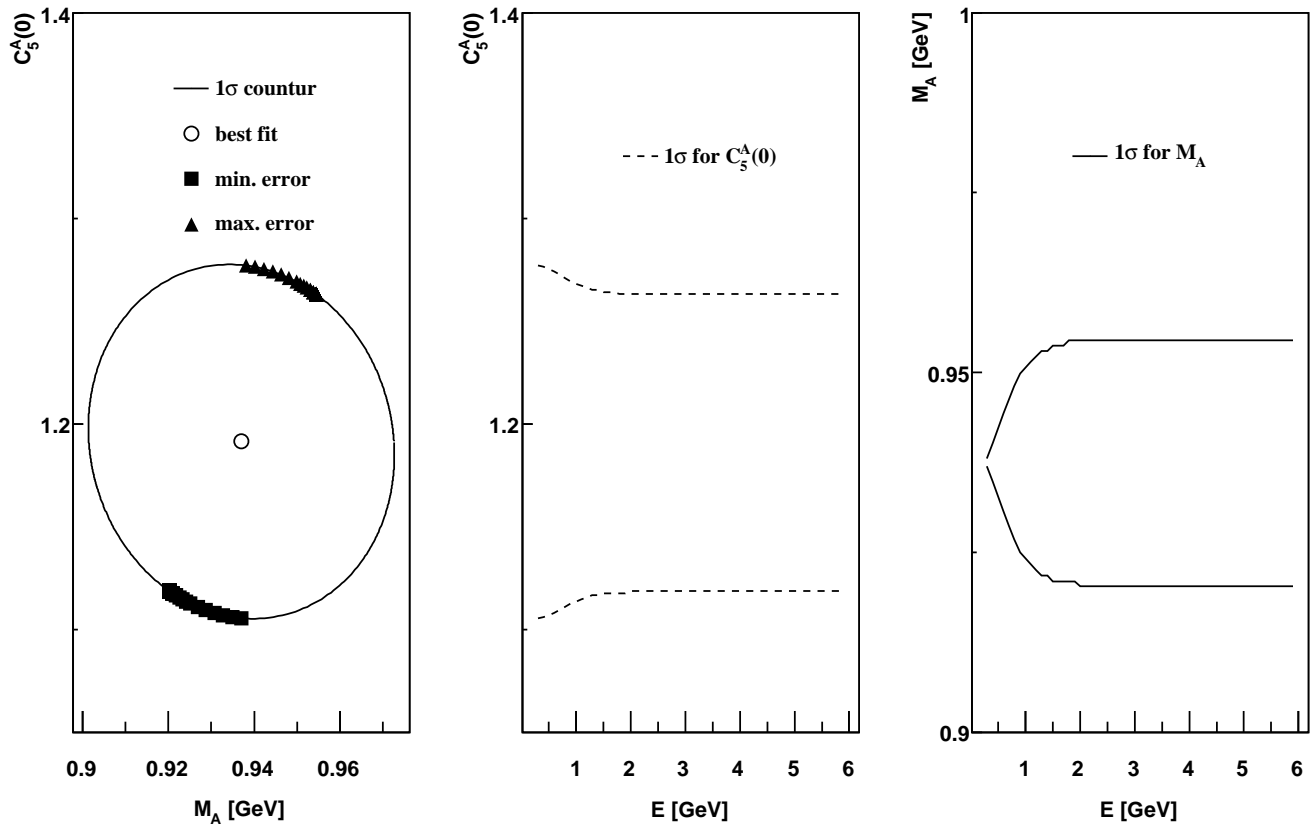}}
\caption{Uncertainties of the fitted parameters. On the left the $1\sigma$ error contour in the $M_A$ and $C_5^A(0)$ plane is plotted. The black squares and triangles denote the points on the contour with maximal and minimal errors for total cross-section for $\nu+p\to \mu^-+p+\pi^+$ at different values of neutrino energy $E$. The obtained points are then mapped in the form of $(C_5^A(0))_{min, max}$ and $(M_A)_{min,max}$ dependence on $E$ (middle and right figures).
\label{fig_c5a_energy}}
\end{figure*}
\begin{figure*}[t!]
\centering{
\includegraphics[width=1.0\textwidth]{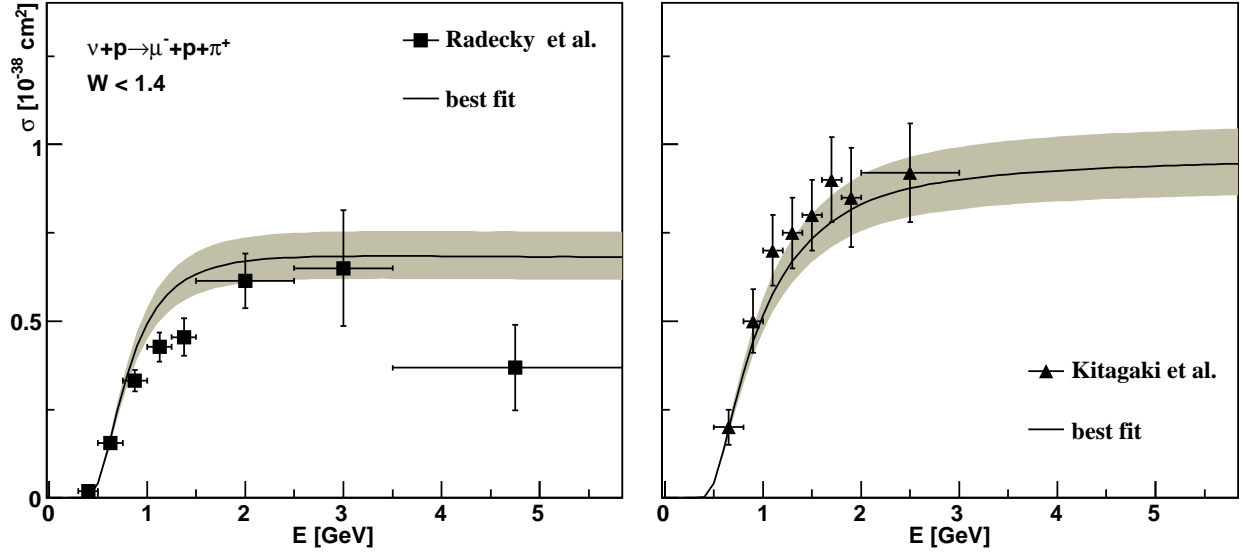}}
\caption{
Total cross-section for $\nu+p\to \mu^- + p+\pi^+$. In the left panel the ANL data \cite{Radecky:1981fn} with the cut $W=1.4$ are shown (black squares), while the right panel presents the BNL data \cite{Kitagaki:1986ct} (without cut in $W$) -- black triangles. The overall normalization error is not plotted.
The best fit curves were obtained with a corresponding cut in $W$. The theoretical curves  were obtained with dipole parametrization Eq. (\ref{c5a_dipol}) with $M_A=0.94$~GeV and $C_5^A(0)=1.19$.  The shaded areas denote the $1\sigma$ uncertainties of the best fit. The theoretical curves are not modified by the deuteron correction effect.
\label{fig_total}}
\end{figure*}

\begin{figure}[t!]
\centering{
\includegraphics[width=0.5\textwidth]{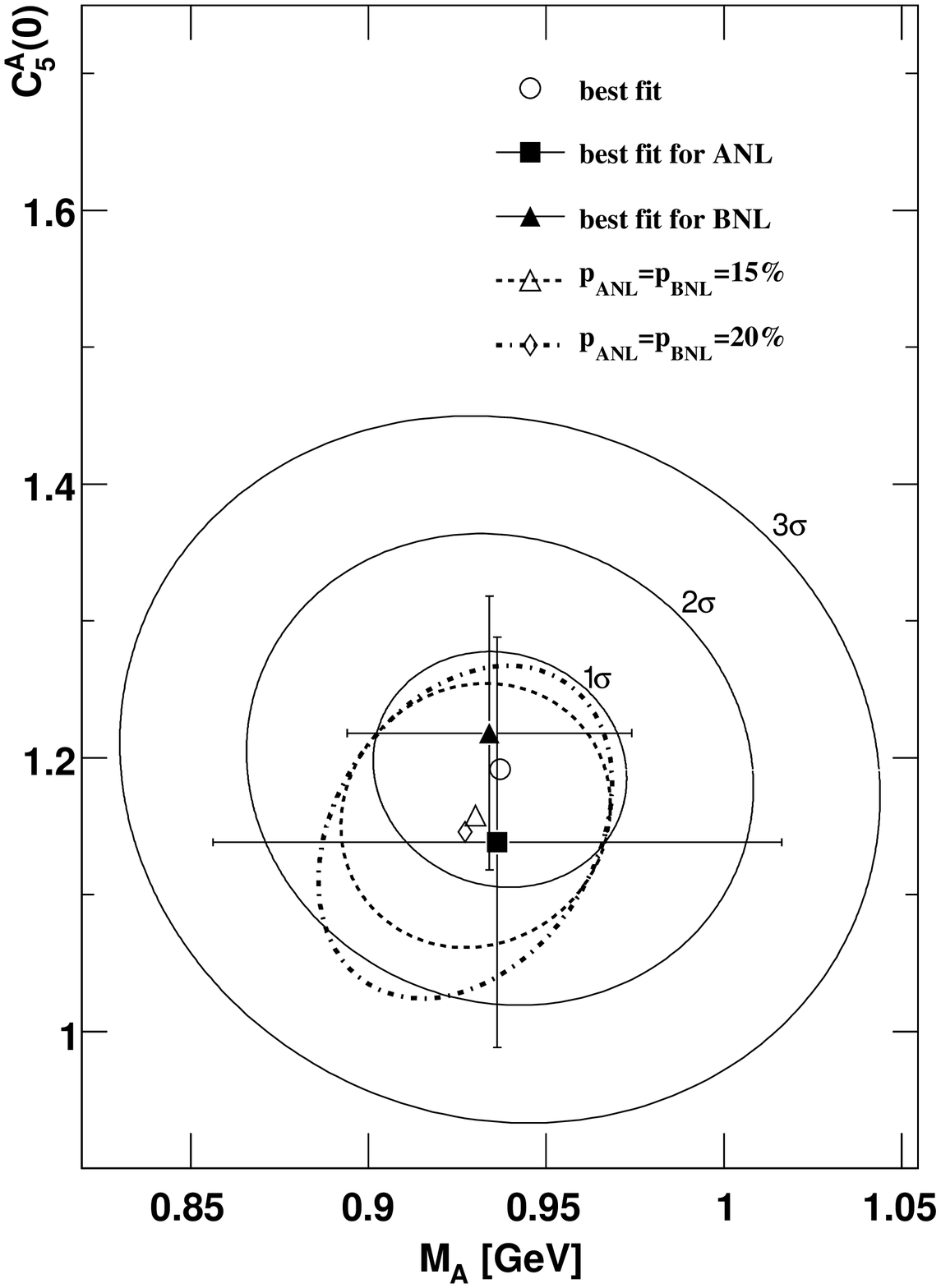}}
\caption{
The $1\sigma$ (solid line), $2\sigma$ (dotted line), and $3\sigma$ (dashed line) error contours
for the $C_5^A(0)$ and $M_A$ fit parameters [see Eq. \ref{c5a_dipol}].
The best fit is denoted by the open circle. The best fit obtained for only ANL/BNL data is represented by black square/triangle.  The open triangle denotes the best fit obtained with $p_{ANL}=p_{BNL}=15\%$ (the $1\sigma$ contour is plotted with a dashed line). The open diamond denotes the best fit obtained with $p_{ANL}=p_{BNL}=20\%$ (the $1\sigma$ contour is plotted with the dashed-dotted line).
\label{fig_c5a_and_ma}
}
\end{figure}

A natural question arises as to why the ANL collaboration did not introduce a similar
$Q^2$ cut on their data. A possible explanation is that in their case
a $Q^2>0.1$~GeV$^2$ cut would eliminate as many as $\sim$ 15\% of the events. This is because the
ANL neutrino beam is of lower energy.

Since the BNL data are given in a different form than in the case of ANL it is natural to use the  expression (\ref{chi2_Ni}).

The theoretical numbers of events in a given $Q^2$ bin are computed with the following formula:
\begin{eqnarray}
N_{th,i}^{BNL}  =  N_{BNL}^{th} \frac{ \Delta Q^2_i \sigma_{th}^{BNL,diff}(Q^2_i)}{\sigma_{tot-th}^{BNL}},
\end{eqnarray}
with
\begin{widetext}
\begin{eqnarray}
\sigma_{th}^{BNL,diff}(Q^2_i) & \equiv &
\frac{1}{\Psi_{BNL}}\cdot\frac{1}{\Delta Q^2_i}\int_{Q^2_i - \Delta
Q^2_i/2}^{Q^2_i + \Delta Q^2_i/2} d Q^2 \int_{E_{min}}^{E_{max}}
\Phi_{BNL}(E)  d E
\int_{M+m_\pi}^{1.4~GeV}dW \sigma_{th}(E, Q^2,W), \\
\sigma_{tot-th}^{BNL} & \equiv &  \frac{1}{\Psi_{BNL}}\int d Q^2
\int_{E_{min}}^{E_{max}} d E
\Phi_{BNL}(E)\int_{M+m_\pi}^{1.4~GeV}dW \sigma_{th}(E, Q^2, W).
\end{eqnarray}
In the case of BNL data the total cross-section is given without any cut on $W$ and we need:
\begin{eqnarray}
\label{sigma_TOT-EX_BNL} \sigma^{exp}_{BNL} &=& \frac{1}{\Psi_{BNL}
}\int_{E_{min}}^{E_{max}} \Phi_{BNL} (E)\sigma_{exp} (E) dE =
0.66\times 10^{-38}\mathrm{cm}^2, \\
\sigma^{th}_{BNL} &=&
\frac{1}{\Psi_{BNL}}\int d Q^2 \int_{E_{min}}^{E_{max}} d E
\Phi_{ANL}(E) \sigma_{th}(E,
Q^2).
\end{eqnarray}
\end{widetext}

\section{Results and discussion}

We analyze simultaneously both the ANL and BNL data. The number of
degrees of freedom is
\begin{equation}
NDF = n_{ANL} + n_{BNL}   - n_{par} - 2,
\end{equation}
where $n_{par}$ is the number of parameters in the analytical expression for the $C_5^A(Q^2)$. The extra factor of 2 comes from two renormalization constants:
$p_{ANL}$ and $p_{BNL}$.

As explained in the Introduction, two parametrization of $C_5^A(Q^2)$ are studied:
\begin{enumerate}
\item[(i)] dipole [Eq. (\ref{c5a_dipol})],
\item[(ii)] Adler [Eq. (\ref{c5a_adler})].
\end{enumerate}

We start the numerical analysis with discussion of the simplest parametrization of the axial form factor, namely the dipole one (\ref{c5a_dipol}).
The results are summarized in Table \ref{table_dipole_results}. There are four different fits: with
$C_5^A(0)=1.15$ or with $C_5^A(0)$ treated as a free parameter and in both cases with and without deuteron corrections. All fits have acceptable goodness.

In the case what we consider  the most reliable, i.e. with $C_5^A(0)$ treated as a free parameter and with deuteron effects included we obtain
\begin{equation}
M_A = 0.94 \pm 0.03\,\,\mathrm{GeV},\qquad C_5^A(0)=1.19\pm 0.08.
\end{equation}
It is interesting to see that in all four cases we obtained similar values for the axial mass.

The inclusion of the deuteron effects does not affect much the theoretical parameters: $M_A$ and $C_5^A(0)$.
This is a consequence of the fact that in wide $Q^2$ range the deuteron effects  mainly change the normalization of the $d\sigma/dQ^2$ reducing it by about 10\%, and its impact becomes compensated by the ANL renormalization parameter $p_{ANL}$.  The normalization of the BNL data is not affected by the nuclear correction and the goodness of fit with the deuteron effects included becomes better.

A similar analysis has been repeated for the Adler parametrization of
$C_5^A(Q^2)$. The results are presented in Table \ref{table_adler_results}. As before, there are four different fits: with
$C_5^A(0)=1.15$ or $C_5^A(0)$ treated as a free parameter and in both cases with and without deuteron corrections. All fits have acceptable goodness. Again, the inclusion of deuteron structure correction affects mainly the normalization of the ANL data and leads to modification of $C_5^A(0)$ by $\sim 3\%$.

In the case that we consider  the most reliable i.e. with $C_5^A(0)$ treated as a free parameter and with deuteron effects included we obtain:
\begin{equation}
M_A = 1.29 \pm 0.07\,\,\mathrm{GeV},\qquad C_5^A(0)=1.14\pm 0.08.
\end{equation}

In Fig. \ref{fig_anl_q2} the plots of $d\sigma /dQ^2(\nu_\mu d\rightarrow \mu \Delta^{++})$ are compared with the ANL data. The experimental points are re-normalized by factor $p_{ANL}$ according to the logic of our analysis.  The theoretical predictions are multiplied by $R(Q^2)$ in order to account for nuclear effects.

In Fig. \ref{fig_bnl_q2} the same is done for the BNL data. Above 1.15~GeV$^2$ some bins are joined together, as explained in the introduction.

We analyzed two different parametrizations of the $C_5^A(Q^2)$ form factor, and both led to the values of $C_5^A(0)$ which are compatible with PCAC result (see Fig. \ref{fig_c5a}).

The quark models predict the direct relationship between  nucleon and resonance axial form factors \cite{Schreiner:1973mj}. For example in the Rein-Sehgal model \cite{Rein:1980wg} (an approach commonly applied to neutrino data analysis) the nucleon and $P_{33}(1232)$ axial form factors have the same $Q^2$ dependence (see e.g. \cite{Graczyk:2007bc}), which, in practice, means the same axial mass.  From that point of view it is interesting to notice that the  values obtained by us for the axial mass of the dipole form factor are
very similar to axial mass parameters extracted from  the quasielastic bubble chamber data in the BNL  \cite{Baker:1981su} ($M_A^{QE}=1.07\pm0.06$~GeV) and in the ANL \cite{Miller:1982qi} ($M_A^{QE}=1.00\pm0.05$~GeV)  experiments.

We examined the impact of a choice of vector form factors (\ref{cv_new}) on the axial mass $M_A$ and on $C_5^A(0)$. We compared the dipole fits obtained by assuming two different parametrizations.  The following numbers were obtained:\\
\\
\begin{tabular}{c|c|c|c|c|c}
 Vector FF  &  $M_A$~(GeV) & $C_5^A(0)$  & $\chi^2$ & $p_{ANL}$ &  $p_{BNL}$ \\
\hline
Equation (\ref{cv_old2}) & 0.93 &  1.24 &\  30.6\  &\ 1.05 &\ 0.97 \\
Equation (\ref{cv_new}) & 0.93 &  1.19 &\  24.3\   &\ 1.08 &\ 0.98
\end{tabular}\\ [0.25cm]
We see that the axial mass does not depend much on the vector form factors which cannot be said about $C_5^A(0)$. For the vector form-factor (\ref{cv_old2}) the higher value of $C_5^A(0)$ is compensated by the smaller $p_{ANL}$ but the shape of the obtained reconstruction of the differential cross-section is worse.

Eventually, we fit the dipole form-factor to the BNL data, but we vary the $Q^2_{cut}$ value.\\ [0.25cm]
\begin{tabular}{c|c|c|c|c|c}
$Q^2_{cut}$ (GeV$^2$) & $M_A$ (GeV) & $C_5^A(0)$ &  $p_{BNL}$& $\chi^2$ & GoF \\
\hline
0.00 & 1.01 & 0.94 &\ 0.96 &\ 40.6\  &\ 0.05 \\
0.05 & 0.96 & 1.14 &\ 0.98 &\ 26.8\  &\ 0.47 \\
0.10 & 0.93 & 1.21 &\ 0.98 &\ 23.3\  &\ 0.67
\end{tabular}\\ [0.5cm]
The dependence of obtained parameters on $Q^2_{cut}$ is significant. As $Q^2_{cut}$ becomes smaller the quality of the fit becomes worse. We note that only BNL data are discussed here and the obtained values for $Q^2_{cut}=0.1$~GeV$^2$ can be recognized as the black triangle in Fig. \ref{fig_c5a_and_ma}.

\subsection{Uncertainties of the fit and statistical consistency of ANL and BNL data }

The uncertainties of our fits and their impact on the uncertainties of cross-sections are analyzed by applying the covariance matrix algorithm \cite{Eadie-71}.  The example of the covariance matrix for the fit with dipole parametrization (with deuteron correction) is given below:
\begin{equation}
V = 10^{-3}\pmatrix{
   1.27 &  -0.23  & 1.19  & 8.2\cdot 10^{-2} \cr
 -0.23 & 7.41 &  7.28   & 0.44 \cr
  1.19  & 7.28   &  0.46   & 0.54 \cr
  8.2\cdot 10^{-2} & 0.44& 0.54 & 0.76}.
\end{equation}
The correlation parameters are
\begin{eqnarray}
\rho_{12} &=& -0.07,\;\; \rho_{13} = 0.33, \;\; \rho_{23} = 0.83, \nonumber \\
\rho_{14} &=& 0.08, \;\;\;\;\; \rho_{24} = 0.19, \;\; \rho_{34} = 0.19,
\end{eqnarray}
where the sequence of the parameters is: $(M_A, C_5^A(0), p_{ANL}, p_{BNL})$.

In Fig. \ref{fig_c5a_energy} the $1\sigma$ error contour in the $M_A$ and $C_5^A(O)$ plane is plotted. The maximal and minimal deviations of the cross-section values from the best fit, due to $1\sigma$ error   are expected to lay somewhere on the error ellipse. For different values of the neutrino energy the points are placed in different region of the ellipse.  We will use these results for the estimation of the uncertainties of the cross-sections obtained with theoretical predictions and the Monte Carlo simulations.

In Fig. \ref{fig_total} the total cross-section for the $\nu+ p \to \mu^+ + p + \pi^+  $ is shown together with $1\sigma$ uncertainties which are shown to be of the order of 10~\%. The theoretical computation (with the dipole axial form factor) is compared with total cross-sections measured at the ANL \cite{Radecky:1981fn} (left panel) and BNL \cite{Kitagaki:1986ct} (right panel) experiments. We present ANL and BNL data on separate plots since the first are obtained with $W_{cut}=1.4$ and the latter with no cut in $W$. The theoretical results are not corrected by the deuteron effect. We do not plot the overall normalization error. The first effect lowers the total cross-sections (especially for the ANL), the second effect enlarges the data error, remembering that, one can say that the fit is compatible with the total cross-section data on the $1\sigma$ level except from one ANL point which is in clear disagreement with all other measurements.

In the case of the global analysis of the data obtained from independent measurements, the classical Pearson $\chi^2$ tests need not work efficiently. In Ref. \cite{Maltoni:2003cu} the statistical test, called parameter goodness of fit, devoted to verification of the statistical compatibility of different measurements is proposed. In the Appendix a short summary of this method is presented. The parameter goodness of fit test measures how far from the global minimum (of $\chi^2_{tot}$) the minima of separate data sets are. Applying this to our analysis we obtain [see Eq. \ref{chi2_pgf_appendix}]
\begin{equation}
\overline{\chi}^2 =0.2, \;\; \mathrm{NDF}_c = 2,
\end{equation}
and the parameter goodness of fit $\mathrm{PGoF} =0.9$ [see Eq. (\ref{PGoF_appendix})]. This means that ANL and BNL data are fully consistent.

To illustrate our results we present in Fig. \ref{fig_c5a_and_ma} the global minimum (denoted by an open circle) with $1\sigma$ error ellipse and  the minima obtained separately for the ANL and BNL data, which are shown by black triangle and square (with $1\sigma$ error bars) respectively. All points lie in the close neighborhood.

\subsection{Impact of flux uncertainties}

One can question whether the $10$\% normalization uncertainty for the BNL neutrino energy spectrum is not too optimistic.

We checked how much our results depend on this assumption and we repeated our computations assuming both uncertainties to be first $15$\% and then $20$\%. For the dipole fit with nuclear effects included, we obtained the following:
\begin{itemize}
\item[(a)] 15\% uncertainties:
\begin{eqnarray}
M_A &=&  0.93\pm 0.04 \;\; \mathrm{GeV},\;\; \nonumber\\
C_5^A(0) &=& 1.16 \pm 0.10,\nonumber\\
p_{ANL} &=& 1.03 \pm 0.11,\nonumber\\
p_{BNL} &=& 0.98 \pm 0.03, \nonumber\\
\label{BNL_w=0.15}
\chi^2/NDF&=& 24.2/28,\;\;  \mathrm{C.L.} =0.67.
\end{eqnarray}

\item[(b)] 20\% uncertainties:
\begin{eqnarray}
M_A &=& 0.93 \pm 0.04\;\; \mathrm{GeV},\;\; \nonumber\\
C_5^A(0) &=& 1.15 \pm 0.12,\nonumber\\
p_{ANL} &=& 1.02 \pm 0.14,\nonumber\\
p_{BNL} &=& 0.98 \pm 0.03, \nonumber\\
\label{BNL_w=0.2}
\chi^2/NDF&=&24.0/28=0.86,\;\;  \mathrm{C.L.} = 0.68.
\end{eqnarray}
\end{itemize}
In Fig. \ref{fig_c5a_and_ma} we added 1$\sigma$ curves for these two cases.

We see that the considered modifications of the normalization errors only very weakly affect the obtained fit for the axial mass. The new central value of $C_5^A(0)$ is in agreement with the one obtained before but the uncertainty becomes much larger.

A general observation is that the increased normalization uncertainty enlarges the uncertainties of the form-factor parameters, and consequently, the cross-section uncertainties. However, we stress that rescaling of the normalization errors  does not destroy the statistical consistency of the ANL and BNL data.

\section{Applications}

\subsection{Evaluation of NC 1$\pi^0$ cross-section uncertainty \label{subsection_pawel}}

It is interesting to investigate what the 1$\sigma$ error contours look like for different $\pi$ production channels in terms of cross-sections.
In particular, one might study two channels that are most important in modern neutrino experiments, i.e.
(i) $\nu_\mu + p \rightarrow \mu^- + \pi^+ + p$, $\nu_\mu + n \rightarrow \mu^- +\pi^+ + n$ (CC $\pi^+$ production)
and (ii) $\nu_\mu + p \rightarrow \nu_\mu + \pi^0 + p$, $\nu_\mu + n \rightarrow \nu_\mu +\pi^0 + n$ (NC $\pi^0$ production). The $\pi^0$ production is the main source of background in  water Cherenkov far detectors of long baseline neutrino experiments searching for $\nu_e$ appearance, like T2K \cite{T2KLOI, Terri:2009zz}. The NC $\pi^0$ events are however difficult to study exclusively; one can try to study them by measuring $\pi^+$ production and extrapolating the results to $\pi^0$ production.

In this section we will use fit results obtained previously to estimate cross-section uncertainty for $\pi$ production channels in the context of T2K experiment. For this purpose two software packages for simulation of neutrino interactions will be used.

The main simulation package used in this analysis was the NuWro Monte Carlo event generator\cite{Juszczak:2005zs}. For resonant $1\pi$ production  this generator uses the Adler-Rarita-Schwinger formalism for the $\Delta$ excitation. The nonresonant part is described by fraction of deep inelastic scattering (DIS) contribution, applying the algorithm described in \cite{Sobczyk:2004va}. NuWro is a generator in which many parameters can be specified manually, including axial form-factor parameters, which are of interest in this analysis. Nuclear effects in oxygen (NEO) have  recently been implemented in NuWro and were used in this study.

As a reference, the Nuance neutrino generator \cite{Casper:2002sd} was also used. Nuance is a widely used tool, tested in experiments with water Cherenkov detectors like K2K and Super-Kamiokande (SK). This appears to be consistent with the measurements of $\pi^0$ production in the 1KT near detector of K2K \cite{Nakayama:2004dp, Mine:2008rt}, as well as atmospheric neutrinos in SK \cite{Ashie:2005ik}. Its implementation of nuclear effects in oxygen can therefore be  considered trustworthy (see Ref. \cite{PawelLadek}).  Resonant pion production in this generator is calculated according to the Rein-Sehgal model \cite{Rein:1980wg}.

We decided to examine the fit that used dipole parametrization of axial form factors in which two parameters $M_A$ and $C_5^A(0)$ were fitted (see Sec. \ref{section_form_factors}). The error ellipse, presented in Fig.\ \ref{fig_c5a_energy}, is calculated for $\Delta^{++}$. However, we can scan this ellipse, calculating NC $\pi^0$ production cross-section for each point on it, and find the minimum and maximum values (which correspond to the 1$\sigma$ range). By doing this for a broad range of incident neutrino energy we can obtain $M_A$ and $C_5^A(0)$ parameters corresponding to the lower and upper 1$\sigma$ bounds at each energy and then use them in NuWro simulations, which will allow us to calculate cross-sections for the channels of interest. The  NuWro simulation package uses exactly the same form factors as the ones that were used in the fit.

All simulation samples in this work were created using water as a target - the most suitable material when simulating interactions in Super-Kamiokande, the far detector of the T2K experiment. Only muon neutrino interactions were taken into account as they dominate the T2K beam \cite{Kudenko:2008ia}.

\begin{figure}[htp]
{\centering
\includegraphics[width=0.45\textwidth,height=0.2\textheight]{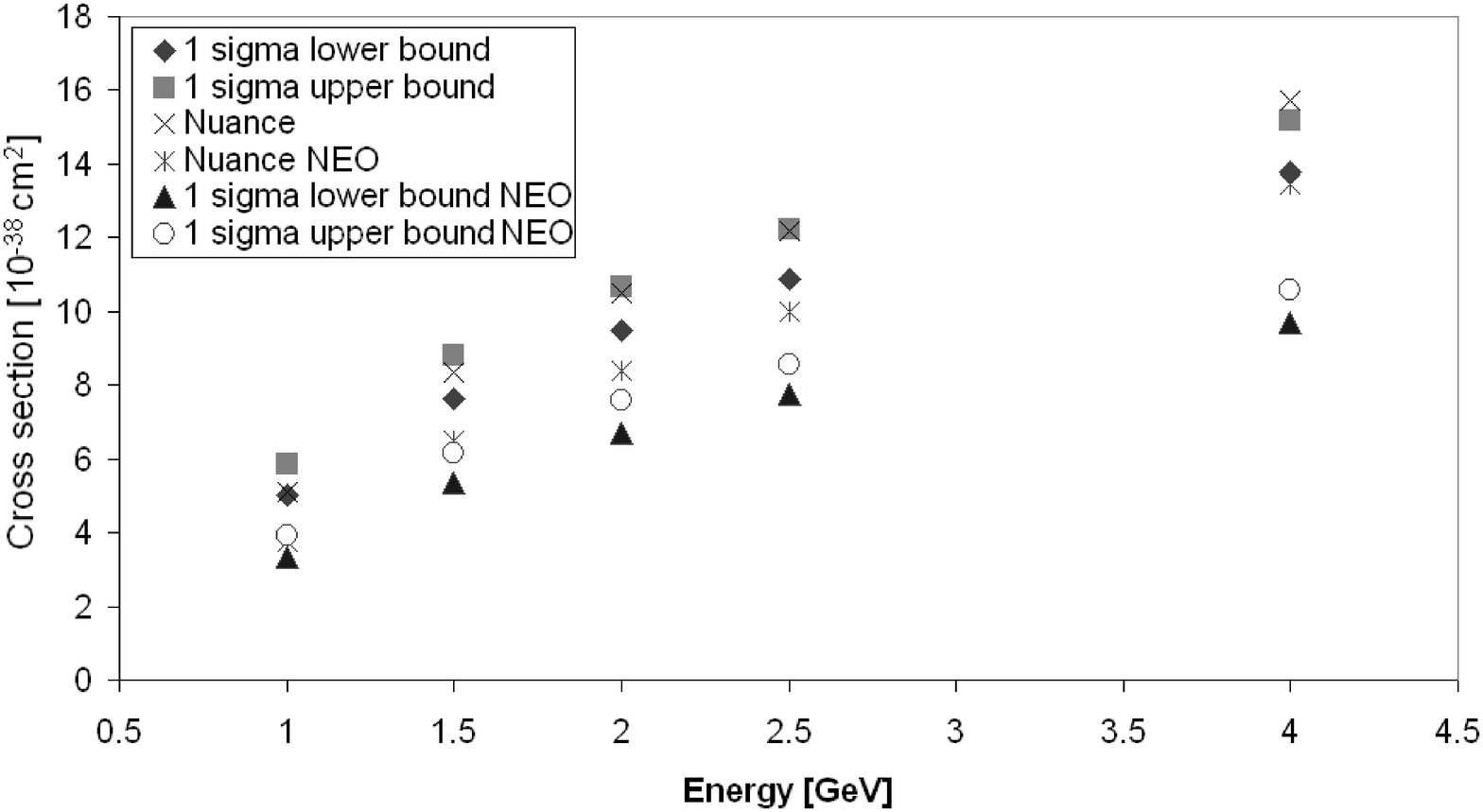} \\[0.5 cm]
\includegraphics[width=0.45\textwidth,height=0.2\textheight]{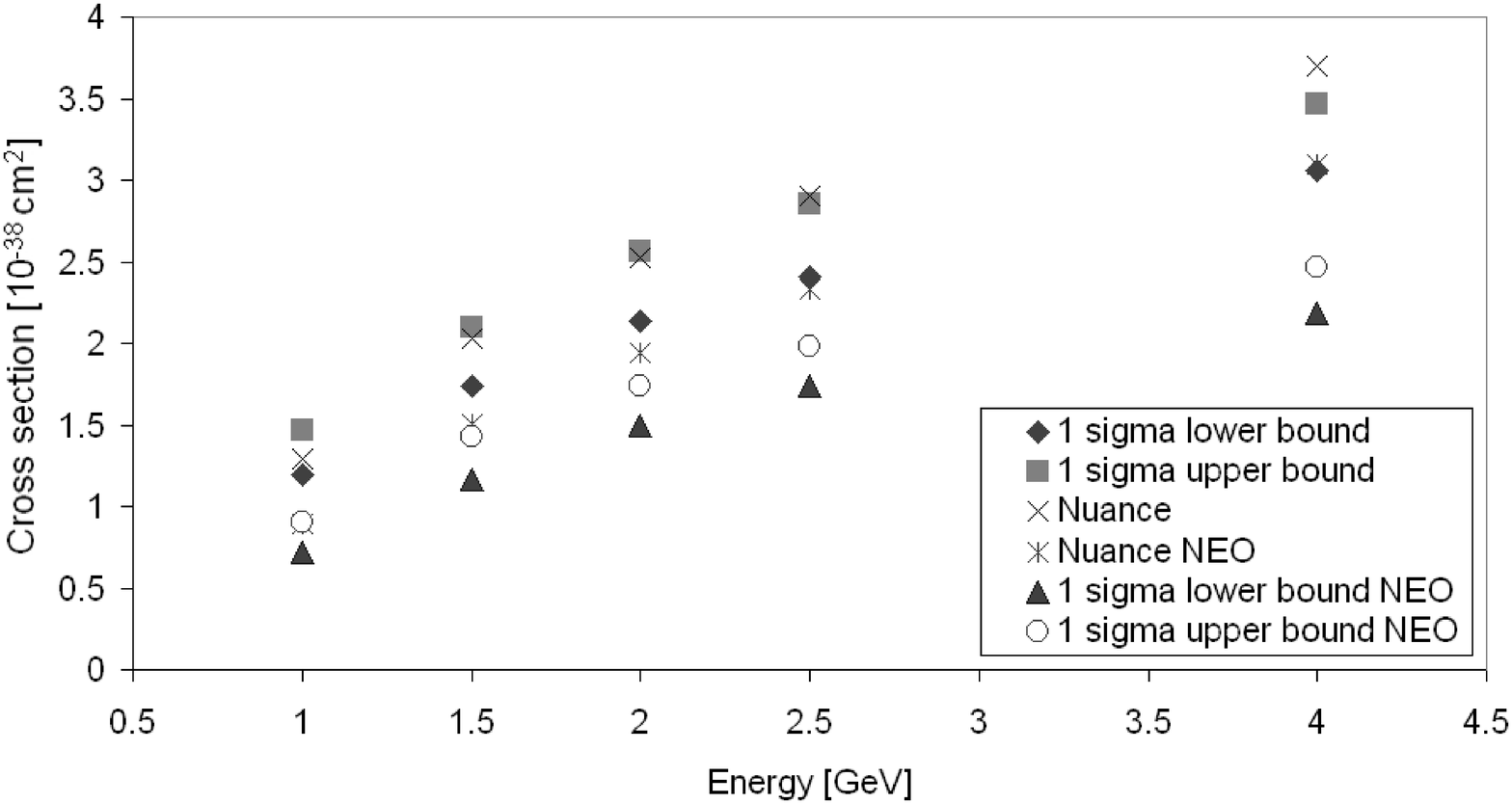} \\[0.4 cm]
\includegraphics[width=0.45\textwidth,height=0.2\textheight]{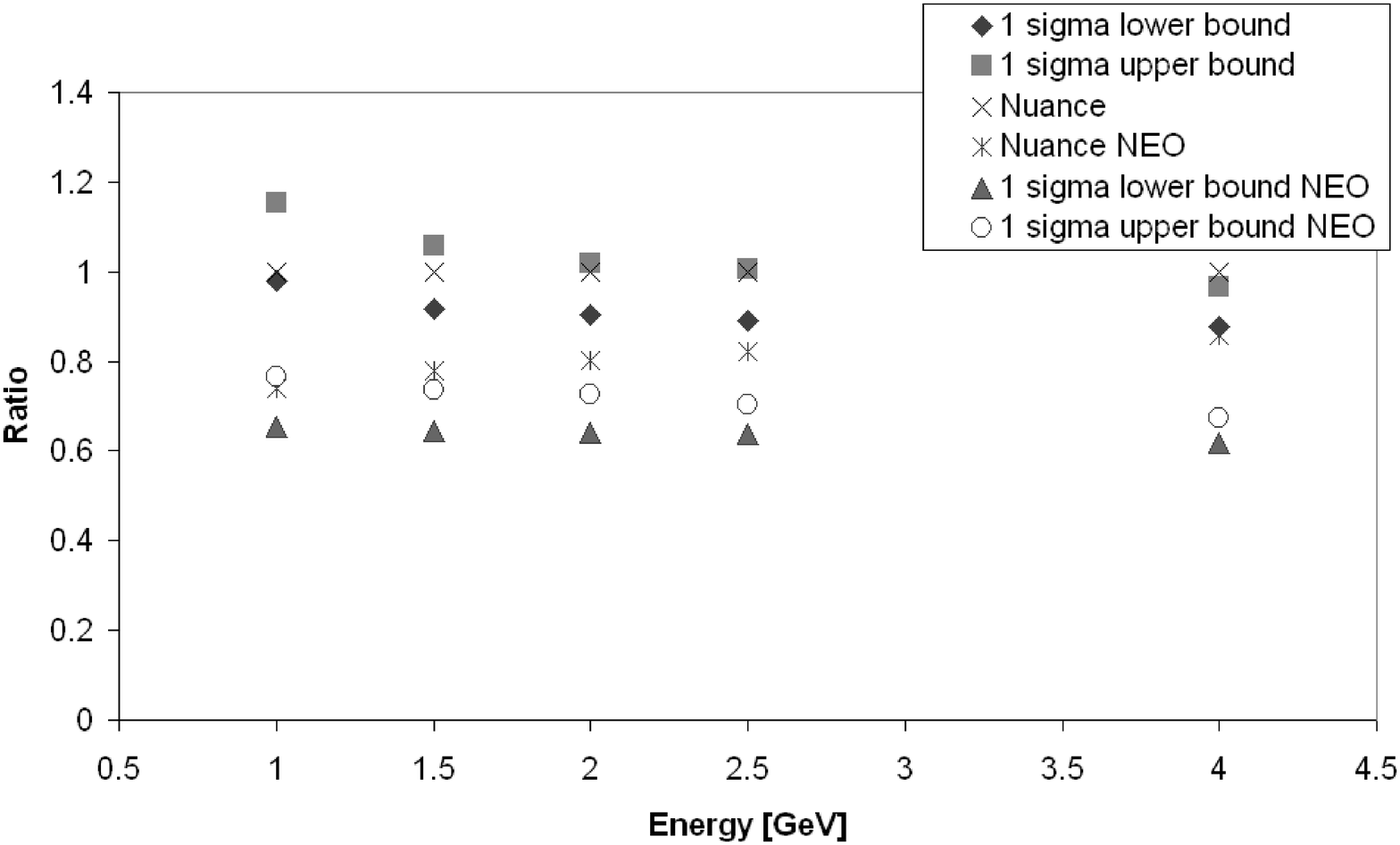} \\[0.4 cm]
\includegraphics[width=0.45\textwidth,height=0.2\textheight]{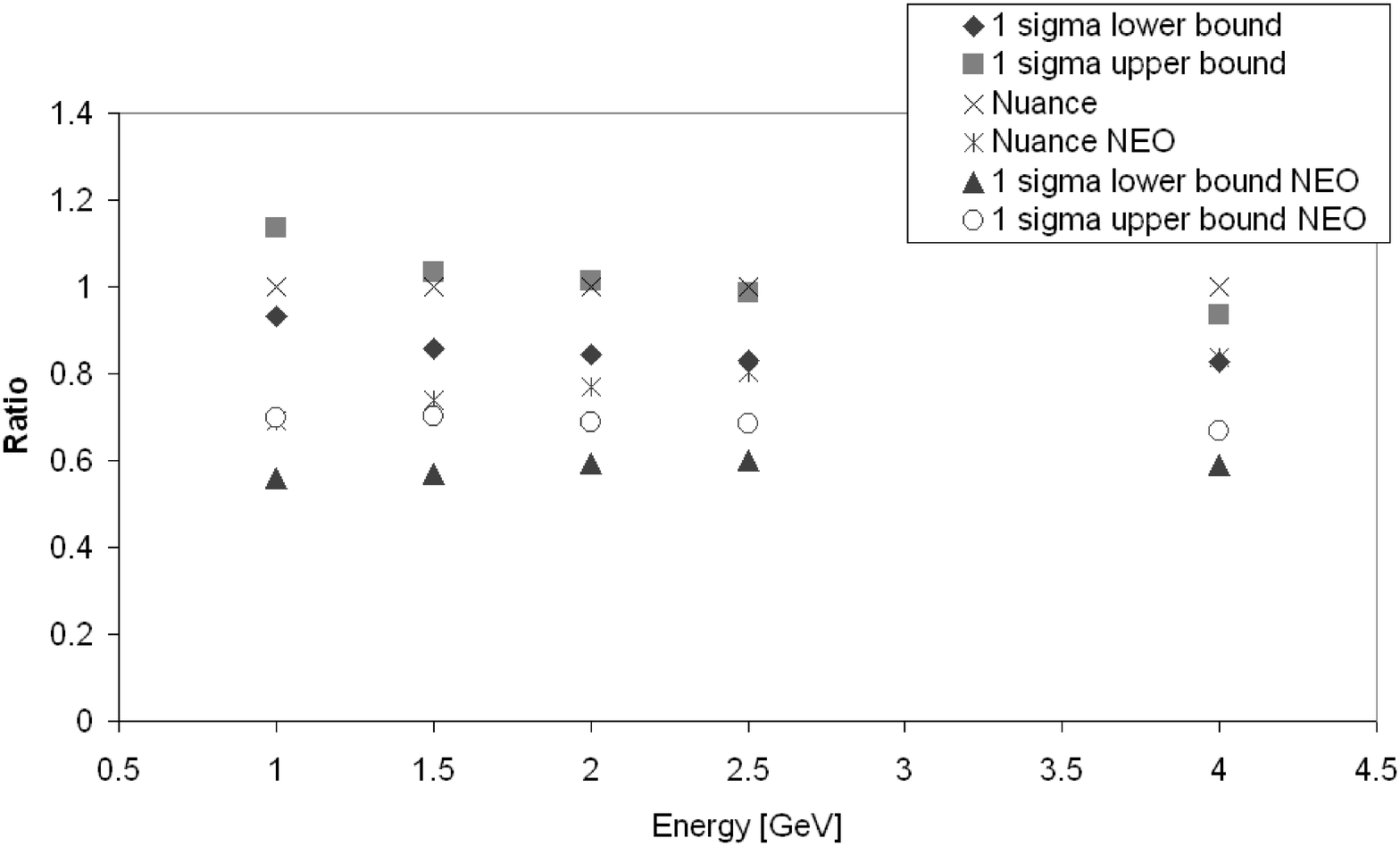}
\caption{cross-sections of $\nu_\mu$ SPP production on water. From top to bottom - CC $1\pi^+$ production absolute cross-section, NC $\pi^0$ production  absolute cross-section, CC $\pi^+$ production cross-section normalized to Nuance results without NEO, and NC $\pi^0$ production cross-section normalized to Nuance results without NEO. NuWro points show 1$\sigma$ error contours. Nuance points are shown here for reference.
\label{fig:ppxsec}
}}
\end{figure}

Figure \ \ref{fig:ppxsec} shows how cross-section uncertainty depends on energy of incoming neutrino. Two sets of points illustrate 1$\sigma$ bounds on cross-sections (cf. Fig.\ \ref{fig_c5a_energy}) obtained using NuWro; others are Nuance results. In order to show how the inclusion of NEO modifies the cross-sections we present separately results with nuclear effects turned off.
All comparisons were done in the two $\pi$ production channels described earlier. Lower plots in Fig.\ \ref{fig:ppxsec} show all cross-sections divided by Nuance results without NEO. It is seen that the differences are most notable in the low energy region. In particular at 1 GeV the uncertainty for $\pi^0$ production is about $\pm 10\%$.

In order to evaluate what neutrino energies are relevant for the T2K $\nu_e$ appearance search, one has to look at the typical energies of pions that constitute background. The oscillation signal from $\nu_\mu \rightarrow \nu_e$ is peaked around 0.7 GeV of the the neutrino energy \cite{T2KLOI}. This means that dangerous $\pi^0$s have energy in this region. Considering energies between 0.4 and 0.9 GeV and using Nuance simulation to translate them into incident neutrino energies, we get neutrino energy range 1 to 4 GeV. We can then conclude that neutrino of such energies are the most important in the context of $\nu_e$ appearance searches in the T2K experiment. Therefore the expected uncertainty varies between 6\% and 10\%. However, it is important to note that the errors discussed here come only from uncertainties in form-factors and do not take into account other model approximations, e.g. nonresonant background in pion production or an influence of matter on the width of $\Delta$ resonance.

We also notice that Nuance predicts higher (with respect to NuWro) SPP cross-sections on nuclear targets. This is probably caused by different assumptions for pion absorption cross-section at higher values of kinetic energy. This is a region where the experimental data for pion nucleus absorption are missing and Monte Carlo predictions have to rely on theoretical assumptions.

\subsection{Comparison with measured CC1$\pi^+$/CCQE ratio}

\begin{figure}[t!]
\centering{
\includegraphics[width=0.5\textwidth]{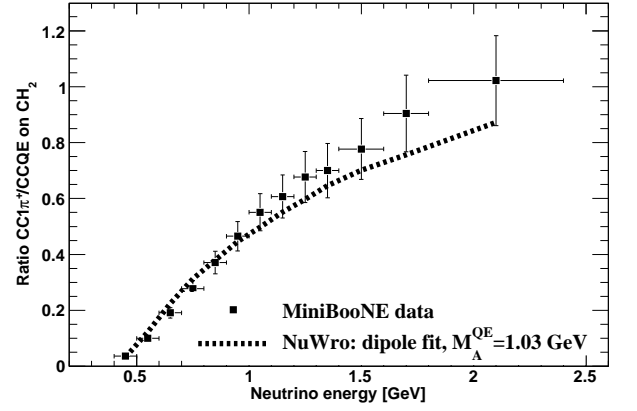}}
\caption{The ratio CC1$\pi^+$/CCQE on the CH$_2$ target. Data points are taken from \cite{MBRatio}. The dipole fit of this paper was used in the NuWro Monte Carlo event generator with QE axial mass $M_A^{QE}=1.03$~GeV.
\label{ratio}
}
\end{figure}

Recently there has been much controversy about the overall normalization of the neutrino cross-section. The MiniBooNE and NOMAD experiments report very different values of the QE axial mass ($M_A=1.35\pm 0.17$~GeV $\kappa =1.07\pm 0.07$ or $M_A=1.37\pm 0.12$~GeV with $\kappa =1.0$ and $M_A=1.05 \pm 0.08$~GeV respectively). For this reason the most reliable  measurements are free from normalization uncertainties, and an interesting example of such measurements is the recent CC1$\pi^+$/CCQE ratio from MiniBooNE. In the Ref. \cite{MBRatio} two sets of data are presented, one for $1\pi^+$ on the nuclear target CH$_2$ and the second with evaluation on the same ratio on isoscalar nuclear target. We compare with the first set of data which is more independent on models contains in MiniBooNE's Monte Carlo generator of events. The value of the QE axial mass used in the simulation was $M_A^{QE}=1.03$~GeV. In Fig. \ref{ratio} we see that we obtained quite good agreement with the MiniBooNE data. The MiniBooNE CC 1$\pi^+$/CCQE data imply that the values of QE and $\Delta$ production the axial masses are {\it correlated}. If one increases in NuWro the value of the axial mass by 30\% as suggested by the MiniBooNE QE data, then in order to keep agreement with the measured ratio, the value of axial mass in $C_5^A(Q^2)$ would have to be accordingly increased.

\section{Final remarks}

We have demonstrated that old bubble chamber ANL and BNL SPP data are self-consistent and restrictive enough to enable one to extract quite precise information about the axial form-factor $C_5^A(Q^2)$. An important ingredient of our reanalysis  was the inclusion of normalization (flux) uncertainty. We also took into account  deuteron nuclear effects but their impact on the final result was surprisingly small due to the interplay with normalization factors.

The evaluation of nuclear effects presented in Ref. \cite{AlvarezRuso:1998hi} was based on nonrelativistic approximation and it would be interesting to perform exact computations and investigate how much this would modify the numerical results of this paper.

It would also be  interesting to try to extract information about the nonresonant background contained in ANL and BNL SPP data on bound neutron with deuteron effects included. Here the situation becomes more complicated because deuteron FSI effects like Pauli blocking can play an important role.

\section*{Acknowledgements}
We thank Luis Alvarez-Ruso for sending us his notes with computations of
deuteron nuclear effects and also the numerical code.

The authors acknowledge support from the Ministry of Science and Higher Education Project  No. DWM/57/T2K/2007. P.~P. acknowledges  support from Grant No. N202 029933.

\appendix

\section{Parameter goodness of fit}
\label{appendix_parameter_goodness_of_fit}
Suppose that $D$ independent data sets are analyzed. Every data set contains $N_r$ ($r=1,2,...,D$) bins (for a review, see Ref. \cite{Maltoni:2003cu}). Let $P$  be the total number of  parameters which are fitted. One can construct the $\chi^2_r$ for the $i$-th data set
Thus the total $\chi^2$ for the global fit reads
\begin{equation}
\label{chi2_total_appendix}
\chi^2_{tot} = \sum_{r=1}^{D} {\chi^2_r}
\end{equation}

The idea of the parameter goodness of fit is to consider a redefined $\overline{\chi}^2$, namely
\begin{equation}
\label{chi2_pgf_appendix}
\overline{\chi}^2 = \chi^2_{tot} - \sum_{r=1}^{D}\chi^2_{r,min},
\end{equation}
where $\chi^2_{r,min} \equiv  min(\chi^2_{r}) $.

It can be shown that $\overline{\chi}^2$ is distributed with:
\begin{equation}
\label{NDF_c_appendix}
\mathrm{NDF}_c = \sum_{r=1}^{D} {P}_r - P.
\end{equation}
degrees of freedom.

Then, the parameter goodness of fit is defined as:
\begin{equation}
\label{PGoF_appendix}
PGoF = \mathrm{CL}(\overline{\chi}, \mathrm{NDF}_c).
\end{equation}
Notice that $\overline{\chi}$ has a minimum at the same point as $\chi^2_{tot}$.

\end{document}